\documentclass{article}
\usepackage{amsmath}

\input{tcilatex}

\begin{document}

\title{\textbf{Liquid-gas phase transition to first order of an argon-like
fluid modeled by the}\\
\textbf{\ hard-core similar Sutherland potential}}
\author{\textbf{Jianxiang Tian}$^{\thanks{%
Email address: jianxiangtian@yahoo.com.cn}}$ \\
\textit{Department of Physics, Dalian University of Technology }\\
\textit{\ Dalian 116024, P.R.China}\\
\textit{Department of Physics, Qufu Normal University}\\
\textit{\ \ Qufu 273165, P.R.China} \and \textbf{Yuanxing Gui} \\
\textit{Department of Physics, Dalian University of Technology }\\
\textit{\ Dalian 116024, P.R.China}}
\maketitle

\begin{abstract}
In this paper, an argon-like canonical system is studied. We introduce five
hypothesis to deal with the total potential of the system. Then the balanced
liquid-gas coexistence phenomenon is analyzed. Good equations of state and
phase diagram are given.

PACS Codes: 05.20.Jj; 05.70.-a; 05.70.Ce; 05.70.Fh.

Keywords: Canonical Partition Function; liquid-gas coexistence.
\end{abstract}

\section{INTRODUCTION}

Much attention has been paid in recent years to the hard core Yukawa (HCY)
potential as a model for the pair interactions of fluids[1]. The liquid
state theories such as the Mean Spherical Approximation (MSA)[2] and the
Self Consistent Ornstein-Zernike Approximation (SCOZA) are proposed. Recent
studies of the HCY fluid can be found in [2,3] and references therein.
Worthwhile similar hard-core Sutherland potential%
\begin{equation}
u_{ij}\text{\textsc{=\{}}%
\begin{array}{c}
+\infty ,r_{ij}<2r_{0}; \\ 
-\frac{B}{r_{ij}^{3\xi }},r_{ij}\geq 2r_{0}.%
\end{array}
\label{1}
\end{equation}%
was less investigated. $i$ and $j$ sign two particles, $r_{ij}$ is their
distance, and $r_{0}$ is interpreted here to be the radius of a hard core. $B
$ is a constant. When $\xi =1$, this potential is Sutherland potential. It
is generally accepted that $\xi >1$. In this paper, we mention out five
hypothesis to deal with the total potential energy of a balanced canonical \
system, on the basis of which we obtain the canonical partition function by
integrating under the help of Eq.(1).

In section 2 of this paper, the equation of state and the chemical potential
are gotten by thermodynamical formulas. So do their reduced forms. Phase
diagrams under five similar hard-core Sutherland potentials are illustrated .

In section 3 of this paper, We conclude that there is no proper similar
hard-core Sutherland potential which can bring completely right
interpretation to the balanced liquid-gas coexistence phenomenon by the
theory in this paper. We find that a similar Sutherland potential with its
exponent $\xi =1.8$ leads to more accurate forecast not only to the critical
coefficient of argon but also to the phase diagram and the relation of
pressure versus temperature.

\section{THEORY}

\subsection{\protect\bigskip The Canonical Partition Function}

The case of a balanced argon-like gases canonical system in 3 dimension
space is considered firstly. The average volume $v$ of each particle is
defined as%
\begin{equation}
v=\frac{V}{N}=\frac{1}{n},  \label{2}
\end{equation}%
where $n$ is the particle number density, $V$ is the volume of the system,
and $N$ is the particle number of the system.

Suppose that volume $v$ is a three dimensional sphere with its radius $r_{D}$%
. Then we have 
\begin{equation}
\frac{4\pi }{3}r_{D}^{3}=\frac{V}{N}=\frac{1}{n}.  \label{3}
\end{equation}%
It is natural that $r_{D}>r_{0}$. They are illustrated clearly in Fig.(1).

As far as the potential energy is considered, we mention out such hypotheses
as

1. \textit{The potential energy} $u_{ij}$ \textit{of two particles with
their distance }$r_{ij}$\ \textit{is described by Eq.(1).}

2. $U=NU_{i}$.

$U$ is the total potential energy of the system and is divided to each
particle equably. $U_{i}$ is called the total potential energy of particle $%
i $, which benefits from other $(N-1)$ particles and is the sum of the
potential energy contributions of $\left( N-1\right) $ particles to it.

3. $U_{i\rightarrow j}=U_{j\rightarrow i}=\frac{1}{2}U_{ij}$.

$U_{j\rightarrow i}$ is the potential energy contribution of particle $j$ to
the total potential energy of particle $i$, and $U_{i\rightarrow j}$ is the
potential energy contribution of particle $i$ to the total potential energy
of particle $j$. The potential energy $U_{ij}$ is divided to these two
particles equably. Thus,%
\begin{equation*}
U_{i}=\left( U_{1\rightarrow i}+U_{2\rightarrow i}+...+U_{\left( i-1\right)
\rightarrow i}+U_{\left( i+1\right) \rightarrow i}+U_{\left( i+2\right)
\rightarrow i}+...+U_{N\rightarrow i}\right)
\end{equation*}%
\begin{equation}
=\frac{1}{2}\left( U_{1i}+U_{2i}+...+U_{\left( i-1\right) i}+U_{\left(
i+1\right) i}+U_{\left( i+2\right) i}+...+U_{Ni}\right) .  \label{4}
\end{equation}%
\bigskip 4. When the thermal fluctuation is omitted: $U_{1\rightarrow
i}+U_{2\rightarrow i}+...+U_{\left( i-1\right) \rightarrow i}+U_{\left(
i+1\right) \rightarrow i}+U_{\left( i+2\right) \rightarrow
i}+...+U_{N\rightarrow i}=\int_{V}u_{j\rightarrow i}d^{3}\overset{%
\rightarrow }{r_{ij}}=\frac{1}{2}\int_{V}u_{ij}d^{3}\overset{\rightarrow }{%
r_{ij}}$

\bigskip $=\frac{1}{2}\int_{r_{1}}^{r_{2}}u_{ij}4\pi r_{ij}^{2}dr_{ij}.$

When the idea that the total potential energy is divided to each particle
equably is accepted, the total potential energy of arbitrary particle $i$
keeps the same in the case we discussed. Our case is a balanced argon-like
gases canonical system. When the thermal fluctuation is omitted, we do think
that the sum of the potential energy contributions of $\left( N-1\right) $
particles to particle $i$ is equal to the potential energy contribution of
particle $j$ to particle $i$ in the universe space of this balanced system. $%
r_{1}$ is the down limit of the integration and $r_{2}$ is the up limit of
the integration.

5. $r_{1}=r_{D}$, $r_{2}=+\infty $.

The down limit is supposed to be\textit{\ }$r_{D}$. And the up limit is
supposed to be $+\infty $. From Eq. (1), we know that $r_{1}$ should take
the value of $r_{0}$, which is the radius of the hard core. $r_{1}=r_{D}$ is
equal to such an idea that integration $I_{1}=\int_{r_{0}}^{r_{D}}u_{ij}4\pi
r_{ij}^{2}dr_{ij}\ll I_{2}=\int_{r_{D}}^{r_{2}}u_{ij}4\pi r_{ij}^{2}dr_{ij}$%
. Hence $I_{1}$ is not taken into account in the calculation of potential. $%
r_{D}>r_{0}$ is illustrated in Fig.(1).

From hypotheses (1-5), we have the result below%
\begin{equation*}
U=NU_{i}=N\left( U_{1\rightarrow i}+U_{2\rightarrow i}+...+U_{\left(
i-1\right) \rightarrow i}+U_{\left( i+1\right) \rightarrow
i}+...+U_{N\rightarrow i}\right)
\end{equation*}%
\begin{equation}
=N\frac{1}{2}\int_{V}u_{ij}d^{3}\overset{\rightarrow }{r_{ij}}.
\end{equation}%
We define%
\begin{equation}
U_{IJ}=\int_{V}u_{ij}d^{3}\overset{\rightarrow }{r_{ij}},
\end{equation}%
and integrate it as follows%
\begin{equation*}
U_{IJ}=\int_{V}u_{ij}d^{3}\overset{\rightarrow }{r_{ij}}
\end{equation*}%
\begin{equation*}
=\int_{r_{D}}^{+\infty }u_{ij}4\pi r_{ij}^{2}dr_{ij}
\end{equation*}%
\begin{equation*}
=\int_{r_{D}}^{+\infty }\left( -\frac{B}{r_{ij}^{3\xi }}\right) 4\pi
r_{ij}^{2}dr_{ij}
\end{equation*}%
\begin{equation*}
=\frac{-4\pi B}{3-3\xi }\left( r_{ij}\right) ^{3-3\xi }|_{r_{D}}^{+\infty }
\end{equation*}%
\begin{equation}
=\frac{4\pi B}{3-3\xi }\left( r_{D}\right) ^{3-3\xi }.
\end{equation}%
Here we applied $r_{D}>r_{0}$, $\xi >1$.

The Hamiltonian $H$ reads%
\begin{equation}
H=U+\sum_{i=1}^{N}\frac{p_{i}^{2}}{2m},
\end{equation}%
where $m$ is the mass of one particle, $p_{i}$ is the momentum of particle $%
i $.

Thus canonical partition function can be solved as%
\begin{equation}
Q=\frac{1}{N!h^{3N}}\int \exp (-\beta H)d^{3N}pd^{3N}r,
\end{equation}%
with $\beta =1/k_{B}T$, $h$ Plank constant. We calculate Eq.(9) as follows%
\begin{equation*}
Q=\frac{1}{N!\lambda ^{3N}}\int \exp (-\beta U)d^{3N}r,
\end{equation*}%
\begin{equation}
=\frac{1}{N!\lambda ^{3N}}z_{N}(T,V)
\end{equation}%
with thermal wavelength $\lambda =\frac{h}{\left( 2\pi mk_{B}T\right) ^{1/2}}
$, $k_{B}$ is the Boltzmann constant and $z_{N}(T,V)$ is called position
partition function normally.

Inputting Eq.(5) to Eq.(10), we get%
\begin{equation*}
Q=\frac{1}{N!\lambda ^{3N}}\int \exp (-\beta U)d^{3N}r
\end{equation*}%
\begin{equation*}
=\frac{1}{N!\lambda ^{3N}}\int \exp (-\beta N\frac{1}{2}\frac{4\pi B}{3-3\xi 
}\left( r_{D}\right) ^{3-3\xi })d^{3N}r
\end{equation*}%
\begin{equation*}
=\frac{1}{N!\lambda ^{3N}}\int \exp (-\beta N\frac{2\pi B}{3-3\xi }(\frac{3}{%
4\pi })^{\left( 1-\xi \right) }n^{\left( -1+\xi \right) })d^{3N}r
\end{equation*}%
\begin{equation*}
=\frac{1}{N!\lambda ^{3N}}\int \exp (\beta B^{\prime }Nn^{\left( -1+\xi
\right) })d^{3N}r
\end{equation*}%
\begin{equation*}
=\frac{1}{N!\lambda ^{3N}}\int \exp (\beta B^{\prime }Nn^{\sigma })d^{3N}r
\end{equation*}%
\begin{equation*}
=\frac{1}{N!\lambda ^{3N}}\exp (\beta B^{\prime }Nn^{\sigma })\int d^{3N}r
\end{equation*}%
\begin{equation}
=\frac{1}{N!\lambda ^{3N}}\exp (\beta B^{\prime }Nn^{\sigma })V_{f}\bigskip
^{N},
\end{equation}%
with \ $\sigma =\xi -1,$ $B^{\prime }=-\frac{2\pi B}{3-3\xi }(\frac{3}{4\pi }%
)^{\left( 1-\xi \right) }$ and $V_{f}$ is the free volume. $n$\textbf{\ }%
\textit{is the particle number density which keeps conservation when a
balanced canonical system is considered.\bigskip }\ And we do not consider
the fluctuation of $n$ in this paper. For the existence of the radius $r_{0}$
of the hard core, each particle excludes a volume $\frac{4\pi r_{0}^{3}}{3}$
and hence for $N$ particles the excluded volume is $\frac{4\pi r_{0}^{3}}{3}%
N.$ The free volume, therefor $V_{f}=V-Nb$ with $b=\frac{4\pi r_{0}^{3}}{3}$%
. Thus Eq. (11) reads

\begin{equation}
Q=\frac{1}{N!\lambda ^{3N}}\exp (-\beta B^{\prime }Nn^{\sigma })\left(
V-Nb\right) ^{N}\bigskip .
\end{equation}

\subsection{The Equation of State}

We can get the equation of state as follows 
\begin{equation*}
P=k_{B}T\left( \frac{\partial \ln Q}{\partial V}\right) _{N,T}
\end{equation*}%
\begin{equation}
=\frac{Nk_{B}T}{V-Nb}-\sigma B^{\prime }n^{\sigma +1}.
\end{equation}%
\bigskip When we choose $\sigma =1$, the equation of state is written as%
\begin{equation}
\left( P+B^{\prime }n^{2}\right) \left( v-b\right) =k_{B}T.
\end{equation}%
This is just the form of the VDW equation of state.

Here we sign the pressure of the gases $P_{1}$, the pressure of the liquids $%
P_{2}$, the chemical potential of the gases $\mu _{1}$, the chemical
potential of the liquids $\mu _{2}$, the critical temperature $T_{c}$, the
critical pressure $P_{c}$, the critical particle number density $n_{c}$,the
particle number density of the gases $n_{1}$, the particle number density of
the liquids $n_{2}$, the reduced temperature $T^{\ast }$, the reduced
particle number density of the gases $n_{1}^{\ast }$, the reduced particle
number density of the liquids $n_{2}^{\ast },$ the reduced \ pressure of the
gases $P_{1}^{\ast }$,the reduced \ pressure of the liquids $P_{2}^{\ast },$
the particle number of the gases $N_{1}$, the particle number of the liquids 
$N_{2}$ , the volume of the gases $V_{1}$, and the volume of the liquids $%
V_{2}$. The relations between them are%
\begin{equation}
T^{\ast }=T/T_{C},
\end{equation}%
\begin{equation}
n_{1}=N_{1}/V_{1},
\end{equation}%
\begin{equation}
n_{2}=N_{2}/V_{2},
\end{equation}%
\begin{equation}
n_{1}^{\ast }=n_{1}/n_{c},
\end{equation}%
\begin{equation}
n_{2}^{\ast }=n_{2}/n_{c},
\end{equation}%
\begin{equation}
P_{1}^{\ast }=P_{1}/P_{c},
\end{equation}%
\begin{equation}
P_{2}^{\ast }=P_{2}/P_{c}.
\end{equation}%
At critical point, the function $P=P(N,V,T)$ has such qualities as

\begin{equation}
\frac{\partial P}{\partial V}|_{T_{c}}=0,
\end{equation}%
\begin{equation}
\frac{\partial ^{2}P}{\partial V^{2}}|_{T_{c}}=0.
\end{equation}%
Thus we get the critical data by solving Eq.(22) and Eq.(23). They are%
\begin{equation}
n_{c}=\frac{\sigma }{(\sigma +2)b},
\end{equation}%
\begin{equation}
k_{B}T_{c}=\sigma (\sigma +1)(\frac{2}{\sigma +2})^{2}B^{\prime
}n_{c}^{\sigma },
\end{equation}%
\begin{equation}
P_{c}=\frac{\sigma ^{2}}{\sigma +2}B^{\prime }n_{c}^{\sigma +1},
\end{equation}%
\begin{equation}
\frac{n_{c}k_{B}T_{c}}{P_{c}}=\frac{4(\sigma +1)}{\sigma (\sigma +2)}.
\end{equation}%
When $\sigma =1$, they read%
\begin{equation}
n_{c}=\frac{1}{3b},
\end{equation}%
\begin{equation}
k_{B}T_{c}=\frac{8B^{\prime }}{27b},
\end{equation}%
\begin{equation}
P_{c}=\frac{B^{\prime }}{27b^{2}},
\end{equation}%
\begin{equation}
C=\frac{n_{c}k_{B}T_{c}}{P_{c}}=8/3.
\end{equation}%
We are very familiar with these results which are just the reduced data from
the VDW equation of state. $C$ is the critical coefficient. When $\sigma
=0.7432$, $C=3.4201.$

Inputting Eq.(24-26) to Eq.(13), we get the reduced equation of state%
\begin{equation}
P^{\ast }=\frac{4n^{\ast }T^{\ast }\left( \sigma +1\right) }{\left( \left(
\sigma +2\right) -n^{\ast }\sigma \right) \sigma }-\frac{\left( \sigma
+2\right) n^{\ast \left( \sigma +1\right) }}{\sigma }.
\end{equation}

\subsection{\protect\bigskip The Chemical Potential}

Here we can solve the chemical potential as follows%
\begin{equation*}
\mu =-k_{B}T\left( \frac{\partial \ln Q}{\partial N}\right) _{V,T}
\end{equation*}%
\begin{equation}
=-B^{\prime }\left( \sigma +1\right) n^{\sigma }+k_{B}T\ln \left( \frac{n}{%
1-nb}\right) +\frac{nk_{B}Tb}{1-nb}-\frac{3k_{B}T}{2}\ln \frac{2\pi mk_{B}T}{%
h^{2}}.
\end{equation}%
Inputting Eq.(24-26) to Eq.(33), we get the critical chemical potential.
Then the reduced chemical potential is calculated to be%
\begin{equation}
\mu ^{\ast }=\frac{A_{1}+A_{2}}{A_{3}},
\end{equation}%
where $A_{1}$, $A_{2}$ and $A_{3}$ are given by%
\begin{equation}
A_{1}=\left[ -n^{\ast \sigma }+\left( \ln \frac{n^{\ast }\sigma }{b(\sigma
+2-n^{\ast }\sigma )}\right) \frac{4\sigma T^{\ast }}{\left( \sigma
+2\right) ^{2}}+\frac{4\sigma ^{2}T^{\ast }n^{\ast }}{(\sigma +2-n^{\ast
}\sigma )\left( \sigma +2\right) ^{2}}\right] ,
\end{equation}%
\begin{equation}
A_{2}=\left[ -\frac{6T^{\ast }\sigma }{\left( \sigma +2\right) ^{2}}\ln 
\frac{8\pi mB^{\prime }T^{\ast }\sigma ^{(\sigma +1)}(\sigma +1)}{b^{\sigma
}\left( \sigma +2\right) ^{\sigma +2}h^{2}}\right] ,
\end{equation}%
\begin{equation}
A_{3}=\left[ -1+\left( \ln \frac{\sigma }{2b}+\frac{\sigma }{2}-\frac{3}{2}%
\ln \frac{8\pi mB^{\prime }\sigma ^{(\sigma +1)}(\sigma +1)}{b^{\sigma
}\left( \sigma +2\right) ^{\sigma +2}h^{2}}\right) \frac{4\sigma }{\left(
\sigma +2\right) ^{2}}\right] .
\end{equation}

\subsection{\protect\bigskip Balanced Liquid-gas Coexistence Canonical System%
}

We consider the case of the balanced liquid-gas coexistence canonical
argon-like system in 3 dimensional space, which is a unit-two phases system.
Here we do think the two phases are described by the same canonical
partition function as Eq.(12). There are three balanced conditions which
must be satisfied when these two phases are balanced. They are thermal
condition%
\begin{equation}
T_{1}=T_{2},
\end{equation}%
dynamic condition%
\begin{equation}
P_{1}=P_{2},
\end{equation}%
and phase condition

\begin{equation}
\mu _{1}=\mu _{2}.
\end{equation}

For Eq.(15-21), these three conditions can be expressed in the reduced unit
as%
\begin{equation}
T_{1}^{\ast }=T_{2}^{\ast },
\end{equation}%
\begin{equation}
P_{1}^{\ast }=P_{2}^{\ast },
\end{equation}%
\begin{equation}
\mu _{1}^{\ast }=\mu _{2}^{\ast }.
\end{equation}%
From Eq.(32), we have

\begin{equation}
P_{1}^{\ast }=\frac{4n_{1}^{\ast }T_{1}^{\ast }\left( \sigma +1\right) }{%
\left( \left( \sigma +2\right) -n_{1}^{\ast }\sigma \right) \sigma }-\frac{%
\left( \sigma +2\right) n_{1}^{\ast \left( \sigma +1\right) }}{\sigma },
\end{equation}%
\begin{equation}
P_{2}^{\ast }=\frac{4n_{2}^{\ast }T_{2}^{\ast }\left( \sigma +1\right) }{%
\left( \left( \sigma +2\right) -n_{2}^{\ast }\sigma \right) \sigma }-\frac{%
\left( \sigma +2\right) n_{2}^{\ast \left( \sigma +1\right) }}{\sigma }.
\end{equation}

From Eq.(34), we have%
\begin{equation}
\mu _{1}^{\ast }=\frac{\left( A_{1}\right) _{1}+\left( A_{2}\right) _{1}}{%
\left( A_{3}\right) _{1}},
\end{equation}%
\begin{equation}
\mu _{2}^{\ast }=\frac{\left( A_{1}\right) _{2}+\left( A_{2}\right) _{2}}{%
\left( A_{3}\right) _{2}}.
\end{equation}%
The solution of Eq.(41) and Eq.(42) is%
\begin{equation}
T_{1}^{\ast }=T_{2}^{\ast }=T^{\ast }=\frac{\left( (n_{1}^{\ast })^{(\sigma
+1)}-(n_{2}^{\ast })^{(\sigma +1)}\right) \left( 1-n_{c}(n_{1}^{\ast
})b\right) \left( 1-n_{c}(n_{2}^{\ast })b\right) }{(\sigma +1)(\frac{2}{%
\sigma +2})^{2}\left( (n_{1}^{\ast })-(n_{2}^{\ast })\right) }.
\end{equation}%
The solution of Eq.(41) and Eq.(43) is%
\begin{equation}
T_{1}^{\ast }=T_{2}^{\ast }=T^{\ast }=\frac{\left( (n_{1}^{\ast })^{\sigma
}-(n_{2}^{\ast })^{\sigma }\right) }{\sigma (\frac{2}{\sigma +2})^{2}\left[
\ln \frac{(n_{1}^{\ast })\left( 1-n_{c}(n_{2}^{\ast })b\right) }{%
(n_{2}^{\ast })\left( 1-n_{c}(n_{1}^{\ast })b\right) }+\frac{%
n_{c}(n_{1}^{\ast })b}{\left( 1-n_{c}(n_{1}^{\ast })b\right) }-\frac{%
n_{c}(n_{2}^{\ast })b}{\left( 1-n_{c}(n_{2}^{\ast })b\right) }\right] }.
\end{equation}%
Thus we get a function $f(n_{1}^{\ast },n_{2}^{\ast })=0$ easily from
Eq.(48) and Eq.(49).%
\begin{equation*}
f(n_{1}^{\ast },n_{2}^{\ast })=\frac{\left( (n_{1}^{\ast })^{(\sigma
+1)}-(n_{2}^{\ast })^{(\sigma +1)}\right) \left( 1-n_{c}(n_{1}^{\ast
})b\right) \left( 1-n_{c}(n_{2}^{\ast })b\right) }{(\sigma +1)\left(
(n_{1}^{\ast })-(n_{2}^{\ast })\right) }
\end{equation*}%
\begin{equation}
-\frac{\left( (n_{1}^{\ast })^{\sigma }-(n_{2}^{\ast })^{\sigma }\right) }{%
\sigma \left[ \ln \frac{(n_{1}^{\ast })\left( 1-n_{c}(n_{2}^{\ast })b\right) 
}{(n_{2}^{\ast })\left( 1-n_{c}(n_{1}^{\ast })b\right) }+\frac{%
n_{c}(n_{1}^{\ast })b}{\left( 1-n_{c}(n_{1}^{\ast })b\right) }-\frac{%
n_{c}(n_{2}^{\ast })b}{\left( 1-n_{c}(n_{2}^{\ast })b\right) }\right] }.
\end{equation}%
If $\sigma $ is chosen to be a constant, $n_{2}^{\ast }$ can be gotten in
the way of numerical computation by computer when an arbitrary $n_{1}^{\ast
} $ is fixed. Then $T^{\ast }$ is obtained easily from Eq.(48) or Eq.(49).
And $P^{\ast }$ is solved by Eq.(32). Table.(1) is the theoretic data when $%
\sigma =0.8$. Fig.(2) is the correlation of $n_{1}^{\ast }$ and $n_{2}^{\ast
}$ under different similar hard-core Sutherland potentials signed by
different $\sigma $. Fig.(3) is the phase diagrams corresponding with
different $\sigma $. Fig.(4) is the curve of $\ln P^{\ast }$ versus $%
1/T^{\ast }$. In Ref.[4], we have introduced one method called polynomial
approximation to simulate the relation of two variables. Here we can get the
relation of $\ln P^{\ast }$ versus $1/T^{\ast }$ by this method when $\sigma 
$ is fixed. In Ref.[5], the theoretic relation of $\ln P^{\ast }$ versus $%
1/T^{\ast }$ was given by this method in the case of $\sigma =1$.

In 1945, E.A.Guggenheim collected the data of the balanced liquid-gas
coexistence system from experiments and gave out the correlation of $\rho
_{1}^{\ast }$, $\rho _{2}^{\ast }$ and $T^{\ast }$ by the empirical
equations[5,6] below%
\begin{equation}
\rho _{1}^{\ast }=1+0.75(1-T^{\ast })-1.75(1-T^{\ast })^{1/3},  \label{28}
\end{equation}%
\begin{equation}
\rho _{2}^{\ast }=1+0.75(1-T^{\ast })+1.75(1-T^{\ast })^{1/3}.  \label{29}
\end{equation}%
For that%
\begin{equation}
\rho _{1}^{\ast }=mn_{1}/mn_{c}=n_{1}/n_{c}=n_{1}^{\ast },
\end{equation}%
\begin{equation}
\rho _{2}^{\ast }=mn_{2}/mn_{c}=n_{2}/n_{c}=n_{2}^{\ast },
\end{equation}%
we have%
\begin{equation}
n_{1}^{\ast }=1+0.75(1-T^{\ast })-1.75(1-T^{\ast })^{1/3},
\end{equation}%
\begin{equation}
n_{2}^{\ast }=1+0.75(1-T^{\ast })+1.75(1-T^{\ast })^{1/3}.
\end{equation}%
In Eq.(51-52), $\rho _{1}^{\ast }$ is the reduced density of the gases and $%
\rho _{2}^{\ast }$ is the reduced density of the liquids. As far as argon
system is concerned, the inaccuracy of these two equations is generally only
one or two parts per thousand of $\rho _{2}^{\ast }$ or of $\rho _{1}^{\ast
} $ when $T^{\ast }>0.60T_{c}$[5,6]. So, it is acceptable to consider the
data of $(T^{\ast },n_{1}^{\ast }(T^{\ast }),n_{2}^{\ast }(T^{\ast }))$ from
Eq.(55-56) as the experimental ones in this temperature region[5].

E.A.Guggenheim gave a numerical analytic result of the relation between the
reduced temperature and the reduced pressure from experiments by
equation[5,7]%
\begin{equation}
P_{e}^{\ast }=\exp (5.29-5.31/T^{\ast }),  \label{34}
\end{equation}%
which best fits the experimental data for argon when $T^{\ast }>0.60T_{c}$
except a tiny region near the critical point[5,7]. Thus it is acceptable to
consider the data of $(P_{e}^{\ast },T^{\ast })$ from Eq.(57) as the
experimental ones, too[5]. The data from Eq.(55-57) are illustrated as
experimental data in Fig.(2-4) and Table.(2) to compare with the theoretic
ones deduced above.

\section{CONCLUSIONS AND DISCUSSIONS}

In this paper we advanced five hypothesis to deal with the total potential
energy of a balanced canonical system. Hypothesis 1-4 are easy to be
accepted. The key is hypothesis 5. Actually $r_{1}=r_{D}$, $r_{2}=+\infty $
is a simple approximation to a concrete sytem. In labs, the volume of the
sytem we consider is finite:$r_{2}\neq +\infty .$ Additionally, only when we
consider a static sytem in 3 dimensional space, can the average volume of a
particle be regarded as a sphere. And $r_{D}$ is effective. $r_{1}=r_{D}$ is
only a simple effective approximation.

On the basis of the five hypothesis, we got the canonical partition function
Eq.(12). Thus all the thermodynamic quantities will be solved by Eq.(12).
Following, we analyzed the balanced liquid-gas coexistence canonical
argon-like system. The results are partly illustrated in figures and tables.
Fig.(3-4) indicates that $\sigma $ in the region (0,1) leads to more perfect
forecast to experimental data. Applying numerical calculation by computer,
we find $\sigma =0.8$ is the best among the five values of $\sigma $ chosen
in this paper only when phase diagram and the relation of $\ln P^{\ast }$
versus $1/T^{\ast }$ are considered together. It is clearly illustrated in
Figure.(3-4). Then the similar hard-core Sutherland potential is fixed to be
the form of%
\begin{equation}
u_{ij}\text{\textsc{=\{}}%
\begin{array}{c}
+\infty ,r_{ij}<r_{0}, \\ 
-\frac{B}{r_{ij}^{5.4}},r_{ij}\geq r_{0}.%
\end{array}%
\end{equation}%
In this case, the equation of state is 
\begin{equation}
P=\frac{Nk_{B}T}{V-Nb}-0.8\ast B^{\prime }n^{1.8}.
\end{equation}%
Now the critical coefficient $C$ is equal to $3.2143$. But $\sigma =0.8$ is
not the best when the relation of $n_{1}^{\ast }$ versus $n_{2}^{\ast }$ is
considered. Figure.(2) suggests that $\sigma =0.9$ and $1$ are better.

\ \ \ Now we will ask whether a proper $\sigma $, which can bring completely
right forecast to experiments when the liquid-gas coexistence phenomenon is
considered, exisits or not by the theory in this paper. Suppose that such a $%
\sigma =\sigma _{0}$ exists. Thus Eq.(50) will be right when arbitrary two
experimental data terms $\left( n_{1}^{\ast },n_{2}^{\ast }\right) $ are
considered. From Eq.(55-56), we get two experimental data terms:$\left(
0.1963,2.0137\right) |_{T^{\ast }=0.8600}$, $\left( 0.3928,1.6822\right)
|_{T^{\ast }=0.9500}.$ We input them to Eq.(50) and get the numerical value
of $\sigma $ by Matlab software. The result is $\sigma |_{T^{\ast
}=0.8600}=0.9872\neq \sigma |_{T^{\ast }=0.9500}=1.0883.$ Thus we conclude
such a proper $\sigma $ does not exist. But it does not mean that there
exists no value for $\sigma $ corresponding with the potential explaining
the experimental data properly. A new theory may be do. Actually, when the
liquid-gas phase transiton to second order is considered, our work does not
work well for the obvious fluctuation near to the critical point, which is
ommitted in hypothesis (4). We will discuss it in details in future work. Of
course, we see that this work offered the proper critical coefficient $%
C=3.4201$ of argon by a simple equation of state with $\sigma =0.7432$.

\section{ACKNOWLEDGEMENT}

This project is supported by the National Natural Science Foundation of
China under Grant No. 10275008. We thank Mrs.Cuihua Zhang, who is working in
the office of civil administration Sishui, Jining, Shandong Province,
P.R.China, for her help with this paper. We thank Prof. Yunjie Xia and Dr.
Hua Jiang for their recent help with the English expression in this
manuscript.

\section{REFERENCES}

\begin{enumerate}
\item Y. Rosenfeld,\textit{\ J. Chem. Phys}. 98, 8126 (1993).

\item C. Caccamo, G. Giunta and G. Malescio, \textit{Mol. Phys.} 84, 125
(1995).

\item D. Pini, G. Stell and N. B. Wilding, \textit{Mol. Phys}. 95, 483
(1998).

\item Jianxiang Tian, Yuanxing Gui, Guanghai Guo, Yan Lv,Suhong Zhang, and
Wei Wang, \textit{General Relativity and Gravitation}, Vol.35, 1473-1480
(2003), gr-qc/0304009.

\item Jianxiang Tian, Yuanxing Gui, \textit{J.Phas.Equi}, Vol.24, No.6,
533-541 (2003).

\item E.A.Guggenheim, \textit{J.Chem.Phys}, Vol.13, No.7, 253 (1945).

\item E.A.Guggenheim, \textit{Thermodynamics}, North-Holland Physics
Publishing, 138-139 (1967).
\end{enumerate}

\bigskip

\FRAME{ftbpFU}{184.75pt}{175pt}{0pt}{\Qcb{Two radius}}{}{Figure}{\special%
{language "Scientific Word";type "GRAPHIC";maintain-aspect-ratio
TRUE;display "USEDEF";valid_file "T";width 184.75pt;height 175pt;depth
0pt;original-width 181.75pt;original-height 172.0625pt;cropleft "0";croptop
"1";cropright "1";cropbottom "0";tempfilename
'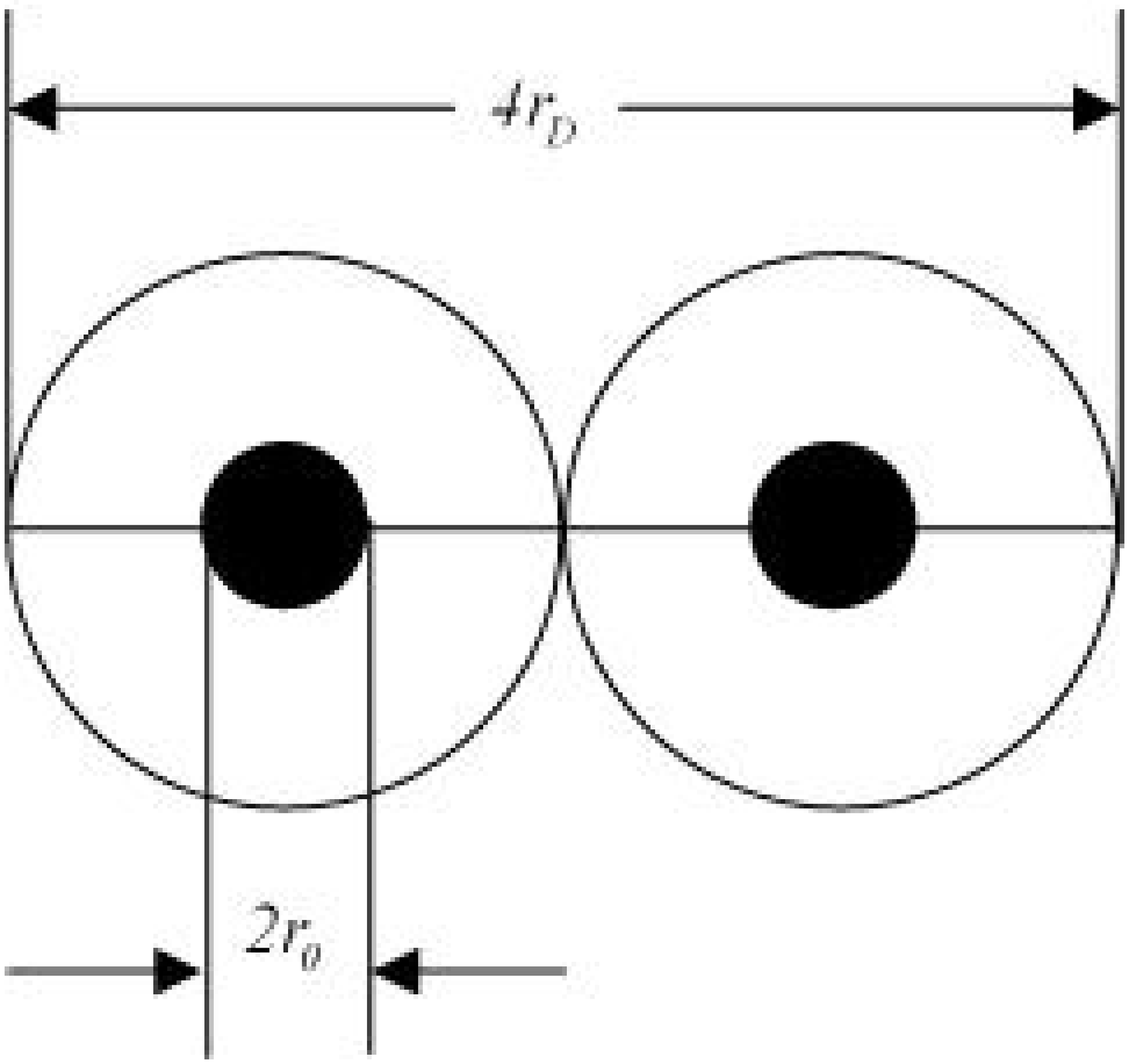';tempfile-properties "XPR";}}

\FRAME{ftbpFU}{339.125pt}{257.375pt}{0pt}{\Qcb{The curve of $n_{1}^{\ast }$
versus $n_{2}^{\ast }$. The real one is from Eq.(60-61). The rest are all
from Eq.(55) with different $\protect\sigma $: square--$\protect\sigma =2$;
diamand---$\protect\sigma =1.5$; star---$\protect\sigma =1$; dot---$\protect%
\sigma =0.9$; cross--$\protect\sigma =0.8$. The star one is the result of
the VDW equation of state. }}{}{Figure}{\special{language "Scientific
Word";type "GRAPHIC";maintain-aspect-ratio TRUE;display "USEDEF";valid_file
"T";width 339.125pt;height 257.375pt;depth 0pt;original-width
513.5pt;original-height 389pt;cropleft "0";croptop "1";cropright
"1";cropbottom "0";tempfilename '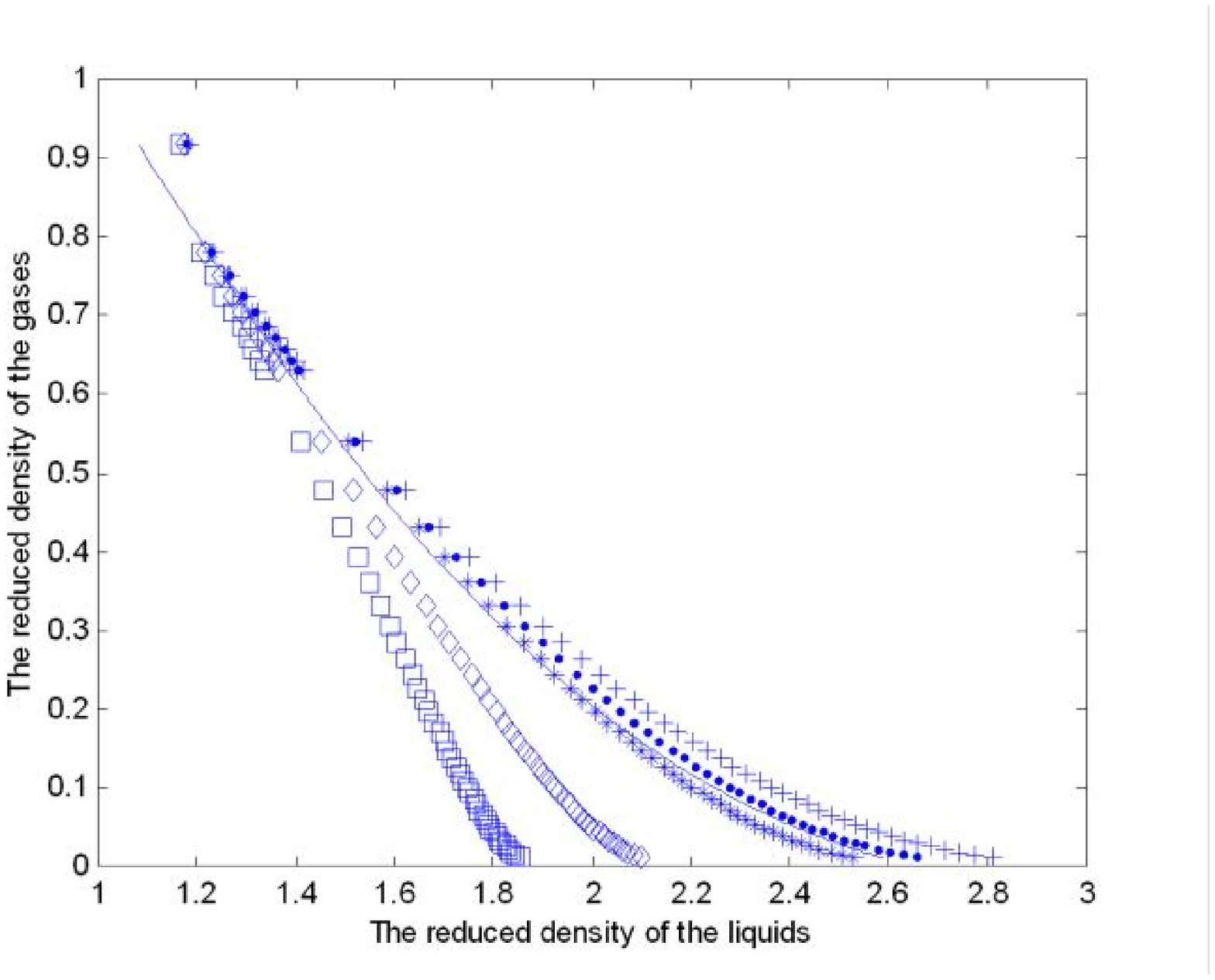';tempfile-properties "XPR";}}

\FRAME{ftbpFU}{314.625pt}{238.8125pt}{0pt}{\Qcb{The phase diagrams. The real
one is from Eq.(60-61). The rest are from the theoretic data with different $%
\protect\sigma $: square--$\protect\sigma =2$; diamand---$\protect\sigma %
=1.5 $; star---$\protect\sigma =1$; dot---$\protect\sigma =0.9$; cross--$%
\protect\sigma =0.8$. The star one is the result of the VDW equation of
state. }}{}{Figure}{\special{language "Scientific Word";type
"GRAPHIC";maintain-aspect-ratio TRUE;display "USEDEF";valid_file "T";width
314.625pt;height 238.8125pt;depth 0pt;original-width 513.5pt;original-height
389pt;cropleft "0";croptop "1";cropright "1";cropbottom "0";tempfilename
'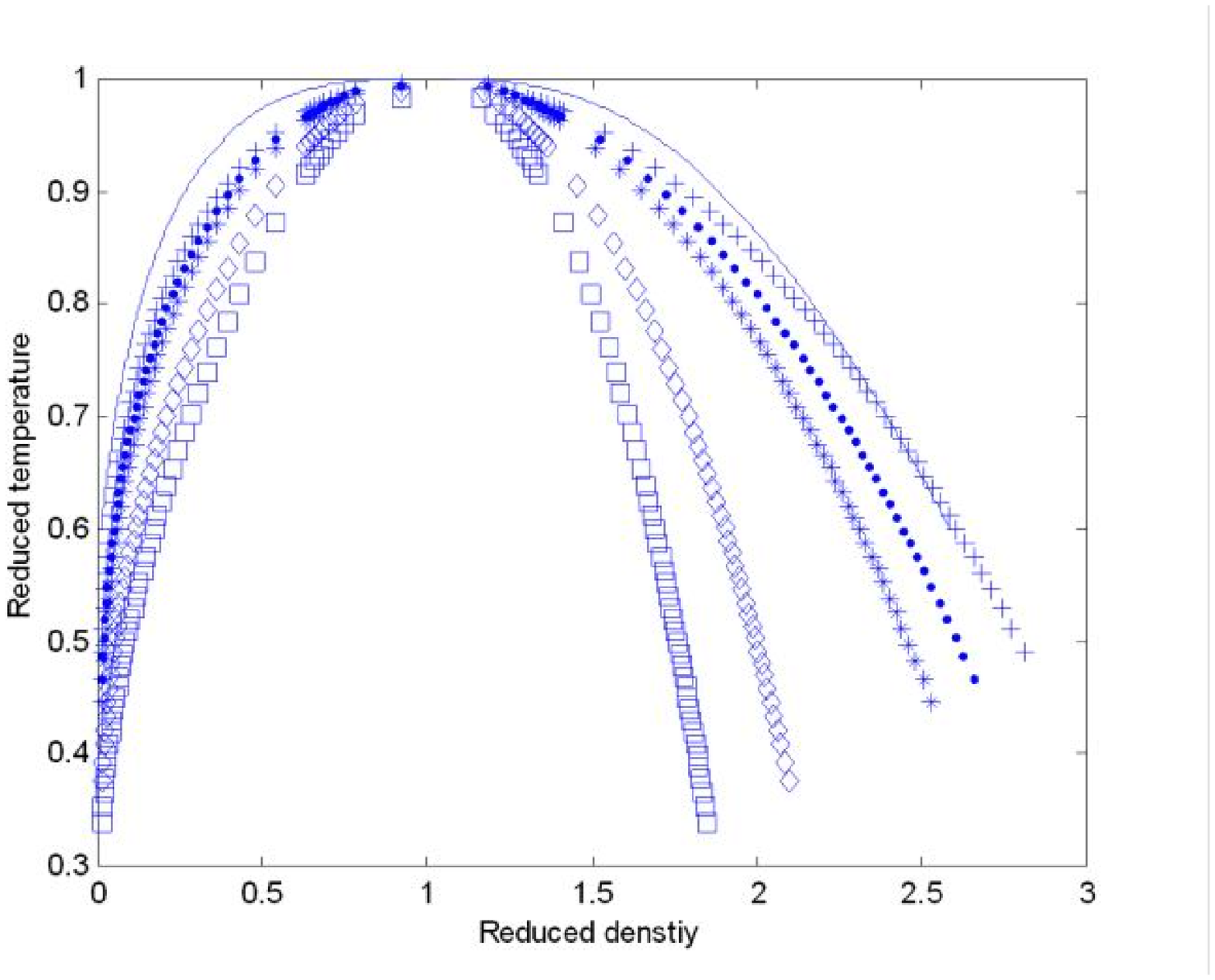';tempfile-properties "XPR";}}

\FRAME{ftbpFU}{288.0625pt}{216.5625pt}{0pt}{\Qcb{The curve of $\ln P^{\ast }$
versus $1/T^{\ast }$. The real one is from Eq.(62). The rest are theoretic
data with different $\protect\sigma $: square--$\protect\sigma =2$;
diamand---$\protect\sigma =1.5$; star---$\protect\sigma =1$; dot---$\protect%
\sigma =0.9$; cross--$\protect\sigma =0.8$. The star one is the result of
the VDW equation of state. }}{}{Figure}{\special{language "Scientific
Word";type "GRAPHIC";maintain-aspect-ratio TRUE;display "USEDEF";valid_file
"T";width 288.0625pt;height 216.5625pt;depth 0pt;original-width
497.9375pt;original-height 373.4375pt;cropleft "0";croptop "1";cropright
"1";cropbottom "0";tempfilename '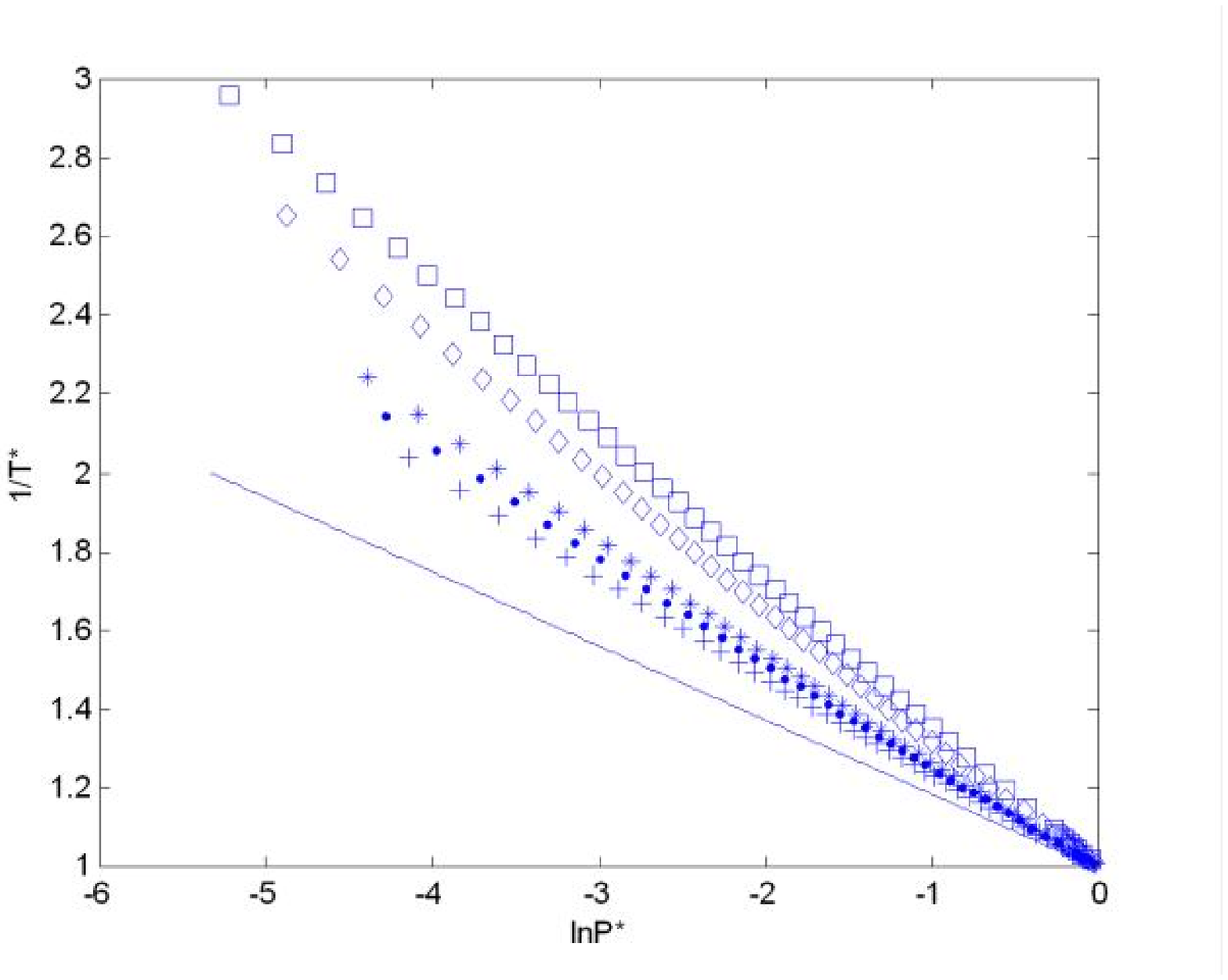';tempfile-properties "XPR";}}

$%
\begin{array}{ccccccccc}
n_{1}^{\ast } & 0.0106 & 0.0292 & 0.0535 & 0.0851 & 0.1266 & 0.1827 & 0.2627
& 0.3928 \\ 
n_{2}^{\ast } & 2.8097 & 2.6574 & 2.5328 & 2.4122 & 2.2860 & 2.1458 & 1.9785
& 1.7521 \\ 
P^{\ast } & 0.0158 & 0.0483 & 0.0931 & 0.1521 & 0.2287 & 0.3280 & 0.4590 & 
0.6411 \\ 
T^{\ast } & 0.4911 & 0.5743 & 0.6360 & 0.6906 & 0.7426 & 0.7944 & 0.8484 & 
0.9085%
\end{array}%
$%
\begin{equation*}
Table.1:\text{Theoretic data when }\sigma =0.8.
\end{equation*}

$%
\begin{array}{ccccccccc}
n_{1}^{\ast } & 0.0106 & 0.0292 & 0.0535 & 0.0851 & 0.1266 & 0.1827 & 0.2627
& 0.3928 \\ 
n_{2}^{\ast } & 2.5894 & 2.4958 & 2.3965 & 2.2899 & 2.1734 & 2.0423 & 1.8873
& 1.6822 \\ 
P_{e}^{\ast } & 0.0284 & 0.0562 & 0.1007 & 0.1670 & 0.2599 & 0.3840 & 0.5434
& 0.7412 \\ 
T^{\ast } & 0.6000 & 0.6500 & 0.7000 & 0.7500 & 0.8000 & 0.8500 & 0.9000 & 
0.9500%
\end{array}%
$

\begin{equation*}
Table.2:\text{The data from Eq.(55-57)}.
\end{equation*}

\end{document}